\documentclass[aps,prl,twocolumn,superscriptaddress,]{revtex4-1}
\usepackage{amsmath}
\usepackage{amsfonts}
\usepackage{graphicx}
\usepackage{color}
\usepackage{dsfont}
\usepackage{verbatim}
\usepackage{mathtools}
\usepackage[normalem]{ulem}
\usepackage{comment}
\usepackage{stackrel}
\usepackage{bm}
\usepackage[pdftex]{hyperref}
\hypersetup{colorlinks = true, urlcolor = black, linkcolor = black, citecolor = black}

\definecolor{myblue}{rgb}{.93, .93, 1}

\newcommand{\bsub}{\begin{subequations}}
	\newcommand{\esub}{\end{subequations}}

\begin{document}
	
\title{Band manipulation and spin texture in interacting moir\'e  helical edges}

\author{Yang-Zhi~Chou}\email{yzchou@umd.edu}
\affiliation{Condensed Matter Theory Center and Joint Quantum Institute, Department of Physics, University of Maryland, College Park, Maryland 20742, USA}

\author{Jennifer Cano}
\affiliation{Department of Physics and Astronomy, Stony Brook University, Stony Brook, New York 11974, USA}
\affiliation{Center for Computational Quantum Physics, Flatiron Institute, New York, New York 10010, USA}

\author{J. H. Pixley}
\affiliation{Department of Physics and Astronomy, Center for Materials Theory, Rutgers University, Piscataway, New Jersey 08854 USA}
\affiliation{Center for Computational Quantum Physics, Flatiron Institute, New York, New York 10010, USA}
 \affiliation{Physics Department, Princeton University, Princeton, New Jersey 08544, USA}

\begin{abstract}
We develop a theory for manipulating the effective band structure of interacting helical edge states realized on the boundary of two-dimensional time-reversal symmetric topological insulators. For sufficiently strong interaction, an interacting edge band gap develops, spontaneously breaking time-reversal symmetry on the edge.
The resulting spin texture, as well as the energy of the time-reversal breaking gaps, can be tuned by an external moir\'e potential (i.e., a superlattice potential). Remarkably, we establish that by tuning the strength and period of the potential, the interacting gaps can be fully suppressed and interacting Dirac points re-emerge.
In addition, nearly flat bands can be created by the moir\'e potential with a sufficiently long period.
Our theory provides an unprecedented way to enhance the coherence length of interacting helical edges by suppressing the interacting gap.
The implications of this finding for ongoing experiments on helical edge states is discussed.
\end{abstract}

\maketitle

\textit{Introduction. --} 
Moir\'e heterostructures, such as magic-angle twisted bilayer graphene, provide tunable platforms to realize novel interaction-driven phases
\cite{tbg1,tbg2,Yankowitz2019,Kerelsky2019,Lu2019,Jiang2019,Xie2019_spectroscopic,Choi2019,Polshyn2019,Cao2020PRL,Polshyn2020,Serlin2020,Sharpe2019,Chen2019signatures,
Po2018TwBLG,xu2018topological,wu2018theory,xie2020nature,zhang2019twisted,lian2019twisted,Wu2019_phonon_PRB,you2019superconductivity,kang2019strong,bultinck2020ground,xie2020topology,lian2020tbg,bernevig2020tbgV,xie2020tbg,bultinck2020mechanism,wang2020chiral,vafek2020towards,chen2020realization,cea2020band,wang2021exact,Wu2020_Collective_PRL,Alavirad2020}. 
Microscopically, these interaction-driven phenomena result from extremely flat bands that are created by the relative twist of two-dimensional (2D) low-energy Dirac cones to a magic angle \cite{Bistritzer,Santos2012}, so that many body interactions become significant. 
Recent proposals have extended this paradigm to other 2D materials \cite{Tang2020simulation,regan20mott,shimazaki2020strongly,wang2019magic,tsai2019correlated,cao2019electric,burg2019correlated,shen2020correlated,liu2020tunable,Pan2020,zang2021hartree}, unconventional superconductors~\cite{can2021high,volkov2020magic}, ultracold atoms in optical lattices~\cite{GonzalezTudela2019,Luo-2021}, and acoustic metamaterials~\cite{gardezi2021simulating}.
Engineering flat bands by tuning to a ``magic angle'' is quite general and may be induced by external quasiperiodic potentials and tunneling processes \cite{Fu2020,SalamonPRL2020,SalamonPRB2020,Chou2020MAS_ch,Fu2020flat} that can be extended to a ``magic-continuum''~\cite{Luoz-2021}.

A moir\'e superlattice potential can also be used to control and manipulate a 2D Dirac cone on the surface of a topological insulator (TI), producing favorable conditions to realize interacting instabilities such as topological superconductivity or quantum anomalous Hall by enhancing the density of states \cite{Cano2021,Wang2021_PRX}.
However, if the surface states are protected by time-reversal symmetry, the formation of a flat, gapped, miniband is topologically obstructed.
Spontaneous symmetry breaking permits an interaction-driven gap, but an analysis of the strong coupling regime is lacking.

\begin{figure}[t]
	\includegraphics[width=0.475\textwidth]{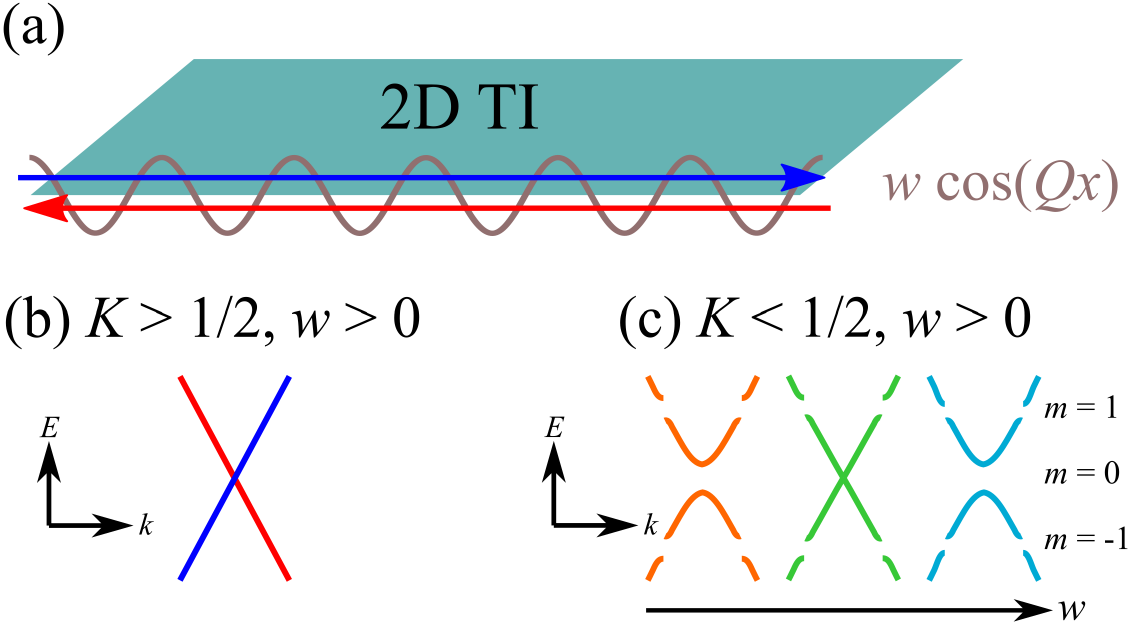}
	\caption{Setup and band structure manipulation. (a) Illustration of the setup. A 2D TI edge state is perturbed by an external periodic potential, $V(x)=w\cos(Qx)$. (b) For a weak repulsive interaction ($1/2<K<1$), the edge band is unaffected by the external potential. (c) For a strong repulsive interaction ($K<1/2$), interaction-induced gaps can open at both zero energy and finite energies. 
	In addition, for particular $w$ and $Q$, the interaction-induced gap may vanish, resulting in a re-emergent gapless edge (middle figure).
	The gaps are labeled by $m$ corresponding to $E=m v Q/4$. 
	}
	\label{Fig:Intro}
\end{figure}

In this Letter, by focusing on the helical edge states of a 2D topological insulator,  we utilize the power of bosonization to provide a clear theoretical demonstration of how strongly coupled interaction-driven phases emerge in moir\'e tuned topological edge states when time-reversal symmetry is spontaneously broken. 
In particular, we study the interplay between interactions and a moir\'e potential (specifically, a periodic potential in the continuum theory) 
on the 1D helical edge states of a 2D time-reversal-invariant topological insulator~\cite{Kane2005_1,Kane2005_2,Bernevig2006,Konig2007,Roth2009,Knez2011,Du2015,Li2015,Du2017,Li2017,Tang2017,Wu2018,Chen2018QSH_WSe2,Ugeda2018WSe2,Reis2017,Stuhler2020}, shown in Fig.~\ref{Fig:Intro}(a). 
Distinct from previous works on the interaction-induced helical spin texture \cite{Sun2015} and RKKY induced orders in the external spins \cite{Hsu2018,Yevtushenko2018},
we focus on the band dispersion of the strongly interacting edges, which can be manipulated by a time-reversal breaking gap at zero energy that develops due to interaction induced backscattering, i.e. umklapp processes \cite{Wu2006,Xu2006} [as sketched in Fig.~\ref{Fig:Intro}(b) and (c)]. 
We show systematically that the spin textures of the interacting gaps can be controlled by both the oscillation period \cite{Sun2015} and the amplitude of the periodic potential and exhibit a nontrivial structure. With the ability to manipulate the gaps in the band structure, we demonstrate that the edge band can be engineered such that re-emergent Dirac points and nearly flat bands can be realized. The re-emergence of the Dirac point at $E=0$ [Fig.~\ref{Fig:Intro}(c)] indicates that the periodic potential can enhance the coherence length of the strongly interacting helical edge state, which is useful for ongoing experiments trying to observe topological insulator edge states in quantum wells~\cite{Du2015,Li2015,Du2017,Li2017,Bubis2021localization}.
Our proposed set up depicted in Fig.~\ref{Fig:Intro}(a) can also be realized in ultracold Fermi gases subjected to a two-dimensional spin orbit coupling~\cite{Huang2016,Song2018}, where the edge potential can be generated via an additional one-dimensional optical lattice.

\textit{Model. --} The edge of a 2D time-reversal-invariant topological insulator is described by a 1D Dirac Hamiltonian perturbed by interaction and an inhomogeneous potential\cite{Kane2005_2,Wu2006,Xu2006}, 
\begin{align}\label{Eq:H_fermion}
\hat{H}=&\sum_{r=R,L}\int dx\left[v_r\psi_r^{\dagger}\left(-i\partial_x \psi_r\right) 
+V(x)\psi_r^{\dagger}(x)\psi_r(x)\right]
\nonumber
\\
+& U\int dx :\left[e^{i\delta Gx}\left(\psi_{L}^{\dagger}\psi_R\right)^2
+\text{H.c.}
\right] + \hat{H}_{\text{LL}}
\end{align}
where $v_r=\pm v_F$ is the Fermi velocity for $r=R/L$
and $\hat{H}_{\text{LL}}$ denotes the forward scattering interactions \cite{Giamarchi_Book} whose precise form is not important here. In the above expressions 
$\psi_R$ ($\psi_L$) is the right-moving (left-moving) fermionic field, $V(x)$ is a spatially varying potential, $U$ encodes the interaction strength, $\delta G=4k_F-G$, $k_F$ is the Fermi wavevector, $G$ is a 
reciprocal lattice vector, and $:\mathcal{O}:$ denotes the normal ordering of $\mathcal{O}$. The right-moving and the left-moving fermions form a Kramers pair.
Under time-reversal, the fermionic fields are transformed as: $\psi_R\rightarrow \psi_L$, $\psi_L\rightarrow -\psi_R$, and $i\rightarrow -i$. 
Thus, time-reversal symmetry forbids elastic single-particle backscattering (e.g., $\psi_L^{\dagger}\psi_R$), but long-wavelength forward scattering (e.g., $\psi_R^{\dagger}\psi_R$) is allowed.
In the presence of Rashba spin orbit coupling, the time-reversal symmetric backscattering interaction [$U$ term in Eq.~(\ref{Eq:H_fermion})] can arise, and spin is no longer a good quantum number.
We focus only on $G=0$ for simplicity, although the results do not rely on this assumption.


Thus, the interacting inhomogeneous helical edge can be described by $\hat{H}=\hat{H}_0+\hat{H}_V+\hat{H}_U+\hat{H}_{\text{LL}}$ \cite{Chou2018,Chou2019}. To study the interplay between interactions and an external periodic potential, we consider
\begin{equation}
    V(x)=w\cos\left(Qx\right),
    \label{eqn:pot}
\end{equation}
where $w$ controls the strength of the potential and $Q>0$ dictates the oscillation in space. We assume that $2\pi/Q$ is not commensurate with the 2D lattice constant $d$.
We also assume that the edge physics does not change the bulk topological phase.

\textit{Commensurate-incommensurate transitions.--} 
The interacting helical fermions can be mapped to an interacting bosonic problem \cite{Shankar_Book,Giamarchi_Book} with 
two bosonic fields ($\phi$ and $\theta$) that are related to  the density and current operators of the original  fermion degrees of freedom  
(see \cite{SM} for a detailed discussion.) 
To incorporate the effect of the external potential, we perform a linear transformation, $\theta'(x)=\theta(x)+\frac{Kw}{vQ}\sin(Qx)$, which yields the bosonized Hamiltonian
\begin{align}
	\nonumber\hat{H}_b'=&\int dx \frac{v}{2\pi}\left[K\left(\partial_x\phi\right)^2+\frac{1}{K}\left(\partial_x\theta'\right)^2\right]\\
	\label{Eq:H_theta'}&\!-\tilde{U}\int\! dx\cos\!\left[4\theta'(x)\!-\!\frac{4Kw}{vQ}\sin(Qx)\!+\!4k_Fx\right],
\end{align}
where $v$ is the velocity of boson, $K$ is the Luttinger parameter, $\tilde{U}=U/(2\pi^2 a^2)$\cite{U_term}, and $a$ is the ultraviolet length scale in the low-energy theory. We focus only on repulsive interactions, corresponding to $K<1$. 
Equation~(\ref{Eq:H_theta'}) is the starting point of our work.

In the absence of the external potential (i.e., $w=0$), the Hamiltonian in Eq.~(\ref{Eq:H_theta'}) maps to the Pokrovsky-Talapov model for the commensurate-incommensurate transition \cite{PokrovskyTalapov,Giamarchi_Book}. At the Dirac point (i.e., $k_F=0$), the model is commensurate, and the cosine term in Eq.~(\ref{Eq:H_theta'}) becomes relevant in the renormalization group sense, inducing a spectral gap for $K<1/2$ \cite{Wu2006,Xu2006,Giamarchi_Book}.
For fillings that are sufficiently away from the Dirac point, the model is incommensurate, and the $\tilde{U}$ term is irrelevant, suggesting a Luttinger liquid behavior.

To study Eq.~(\ref{Eq:H_theta'}) with a nonzero $w$, we express the cosine term ($\tilde{U}$ term) as a Fourier series \cite{FT}:
\begin{align}
\nonumber&\cos\left[4\theta'(x)-\frac{4Kw}{vQ}\sin(Qx)+4k_Fx\right]\\
\label{Eq:cosine_FT}=&\sum_{m=-\infty}^{\infty}|\tilde{F}_m|\cos\left[4\theta'(x)+(4k_F-mQ)x+\chi_m\right],
\end{align}
where $\tilde{F}_m$ corresponds to the Fourier component of $\exp\left[i\frac{4Kw}{vQ}\sin(Qx)\right]$ and $\chi_m$ is the phase of $\tilde{F}_m$. 
Intuitively, Eq.~(\ref{Eq:cosine_FT}) shows that the presence of the potential has reorganized the interaction into a series of backscattering processes with momentum $m Q$, which, in conjunction with the Luttinger parameter, can each open a gap in the excitation spectrum. 
If such a gap, labeled by $m$, opens, then time reversal symmetry on the edge must be spontaneously broken, lifting the topological obstruction to a gapped boundary.
In the following, we investigate Eqs.~(\ref{Eq:H_theta'}) and (\ref{Eq:cosine_FT}) by adopting the analysis in Refs.~\cite{Vidal1999,Vidal2001}, which considered 
a single cosine term ($w=0$ case) \cite{PokrovskyTalapov,Giamarchi_Book}, to the case with multiple cosine terms ($w\neq 0$ case).
As we will show, the gapped edge states exhibit a nontrivial magnetic structure. Further, the magnitude of the gap can be tuned by the strength and period of the potential, allowing a gapless Dirac cone to be realized at special points in the phase diagram.

\begin{figure}[t]
\includegraphics[width=0.475\textwidth]{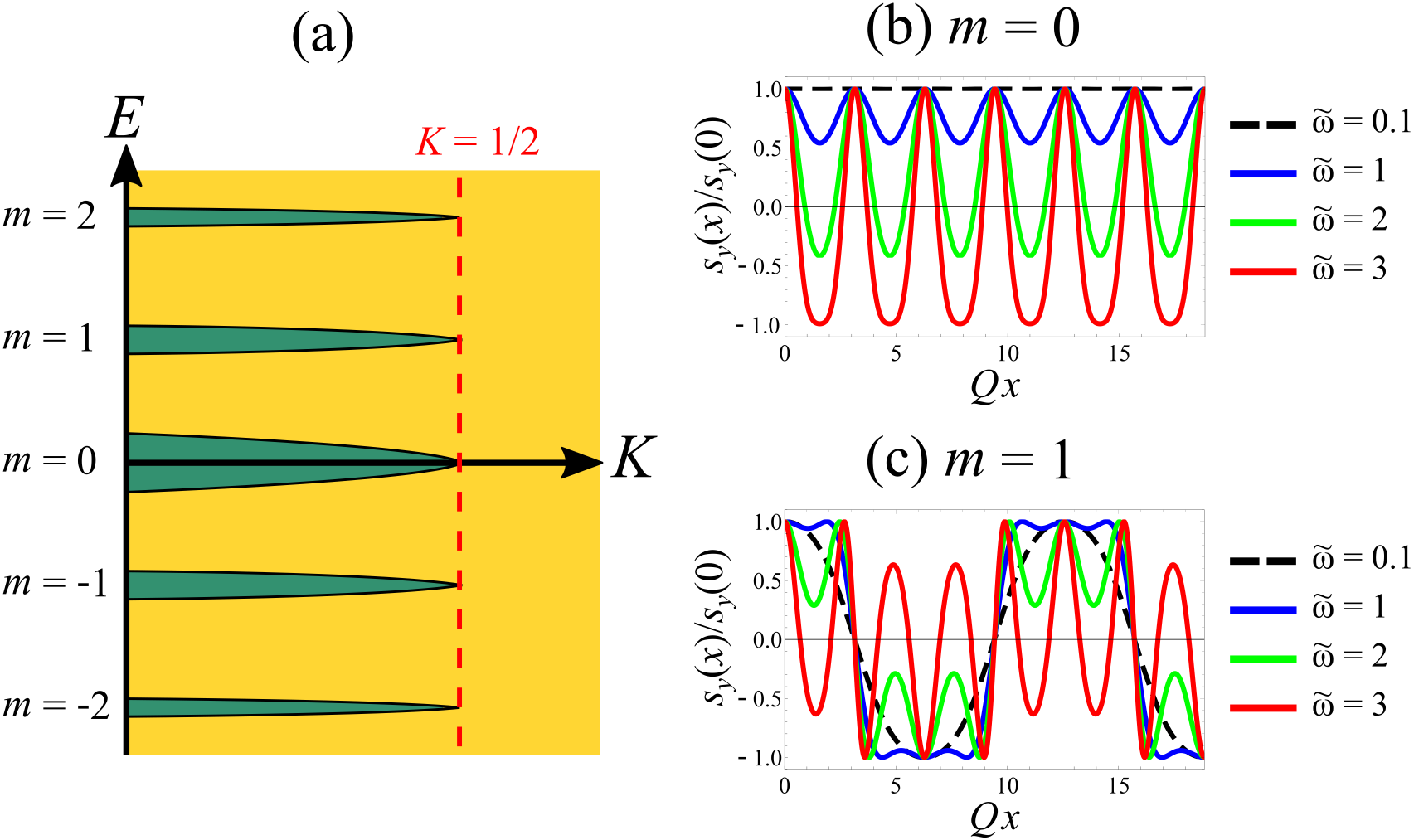}
\caption{Phase diagram and spin texture of the time-reversal breaking order. (a) The phase diagram for $|Q|>4\delta Q_c^{(m)}$ based on bosonization analysis. The yellow region denotes the metallic phase. For $K<1/2$, time-reversal breaking gaps (green regions) can be induced by the interaction. 
The energy of the gap is $E=mvQ/4$ for an integer $m$. The half width of the gap is $v\delta Q_c^{(m)}$, as discussed in the main text. 
(b)-(c): Normalized $s_y(x)$ [Eq.~(\ref{Eq:s_y})] as a function of $Qx$. The phase shift $\zeta_{m,N}$ is set to zero for convenience. (b) $m=0$. (c) $m=1$. We define the dimensionless parameter $\tilde{\omega}=\frac{2Kw}{vQ}$. In both figures, $\tilde{\omega}=0,1,2,3$ correspond to the black dashed lines, blue lines, green lines, and red lines respectively.  
}
	\label{Fig:SDW}
\end{figure}

The commensuration condition for each $m$ [i.e. each cosine term in the summand of Eq.~(\ref{Eq:cosine_FT})] is $4k_F=mQ$.  If this condition is satisfied,
the $\tilde{F}_m$ term in Eq.~(\ref{Eq:cosine_FT}) can open a band gap of 
size
$2v\delta Q_c^{(m)}$ for sufficiently strong interaction ($K<1/2$), as shown in Fig.~\ref{Fig:SDW}(a).
The gapless state will be restored by sufficiently large incommensuration $|4k_F-mQ|>\delta Q_{c}^{(m)}$ as the integral over the cosine term in Eq.~(\ref{Eq:H_theta'}) for this $m$ will vanish.
In general, the threshold of commensuration, $\delta Q_c^{(m)}$, is a complicated function of $\tilde{U}$, $w$ and $K$ \cite{PokrovskyTalapov,Giamarchi_Book}.
If, for each $m$, $Q>4\delta Q_c^{(m)}$, then the interacting gaps in the TI edge spectrum are well separated and we can treat the summand of Eq.~(\ref{Eq:cosine_FT}) as a collection of independent cosine terms.
On the other hand, if $Q\le 4\delta Q_c^{(m)}$, the cosine terms in Eq.~(\ref{Eq:cosine_FT}) interfere with each other; shortly we will employ an exact mapping at $K=1/4$ to study this case.

\textit{Interacting gaps and spin texture. --} For a helical edge state with $K<1/2$, the commensurate cosine term (i.e., satisfying $|4k_F-mQ|<\delta Q_c^{(m)}$) in Eq.~(\ref{Eq:cosine_FT}) becomes relevant, implying the formation of time-reversal breaking gaps at the energies $E(m)=mvQ/4$ 
labelled in Fig.~\ref{Fig:Intro}(c). A phase diagram is illustrated in Fig.~\ref{Fig:SDW}(a).
Since the helical edge state is ``spin-momentum locked,'' the time-reversal breaking gaps carry nontrivial magnetic structures. We discuss the alternative charge picture in \cite{SM}.

To understand the consequence of spin momentum locking on the magnetic texture of each gap
we compute the spin densities $\langle s_x(x) \rangle$ and $\langle s_y(x) \rangle$ (see \cite{SM} for definitions).
%
When the backscattering interaction dominates (i.e., $K<1/2$ and commensurate), 
we employ a mean field approach:
for each mode $m$ in Eq.~(\ref{Eq:cosine_FT}), we solve for $\theta'(x)$ by minimizing the corresponding cosine term and then use
$\theta'(x)=\theta(x)+\frac{Kw}{vQ}\sin(Qx)$ to compute the expectation values of $s_x$ and $s_y$, yielding
\begin{align}
	\label{Eq:s_x}\langle s_x(x)\rangle\approx&\frac{1}{\pi a}\sin\left[\frac{mQx}{2}-\frac{2Kw}{vQ}\sin(Qx)+\zeta_{m,N}\right],\\
	\label{Eq:s_y}\langle s_y(x)\rangle\approx&\frac{1}{\pi a}\cos\left[\frac{mQx}{2}-\frac{2Kw}{vQ}\sin(Qx)+\zeta_{m,N}\right],
\end{align}
where $\zeta_{m,N}=\pi N-\chi_m/2$ is an unimportant phase factor with arbitrary integer $N$.
The expressions in Eqs.~(\ref{Eq:s_x}) and (\ref{Eq:s_y}) reduce to a single frequency oscillation \cite{Sun2015} in the limit $2Kw/(vQ)\ll 1$. Generically, the spin density wave (SDW) states manifest non-sinusoidal oscillations due to the potential in Eq.~\eqref{eqn:pot}. As plotted in Figs.~\ref{Fig:SDW}(b) and (c), the value of $w$ controls the spin texture of the time-reversal breaking order nontrivially, resulting in unusual spin order. In particular, the ferromagnetic order at $E=0$ (equivalently $m=0$) can be deformed to a qualitatively different finite-momentum SDW state in the presence of a sufficiently large $w$.

\textit{Luther-Emery theory and tunable band structure. --} To go beyond the mean field approximation, we turn to the exact Luther-Emery refermionization mapping at $K=1/4$ \cite{Giamarchi_Book}. The interacting bosonized Hamiltonian in Eq.~(\ref{Eq:H_theta'}) maps to a noninteracting massive fermion Hamiltonian:
\begin{equation}
\hat{H}_b\rightarrow \sum_{r=R,L}\int dx\left[\tilde{\Psi}_r^{\dagger}\left(-i\tilde{v}_r\partial_x+\tilde{V}(x)\right)\tilde{\Psi}_r
-\tilde{M}_r\tilde{\Psi}^{\dagger}_{\bar{r}}\tilde{\Psi}_r\right]
\label{Eq:LE_Th}
\end{equation}
where $\tilde{v}_r=\pm v$ for $r=R/L$ is the velocity of the Luther-Emery fermions, $\tilde{V}(x)=V(x)/2+vk_F/2$, $\tilde{M}_r=\pm iM$ for $r=R/L$, $M=U/(2\pi a)$, $\bar{r}$ denotes the opposite direction of $r$,
and the Luther-Emery fermionic field is given by $\tilde{\Psi}_{R/L} = \frac{1}{\sqrt{2\pi a}}e^{i\left[\phi/2 \pm 2\theta\right]}$.
The mass term of the Luther-Emery fermions arises from the backscattering interaction [$U$ term in Eq.~(\ref{Eq:H_fermion})]. 
The Luther-Emery fields ($\Psi_R$ and $\Psi_L$) are not simply related to the physical low-energy fermions ($R$ and $L$). Nevertheless, we can extract information from the Luther-Emery theory such as the energy spectrum, 
density, and current \cite{Chou2018}.

For the Luther-Emery fermions, the band gap induced by spontaneous time-reversal symmetry breaking is $2|M|$. The appearance of finite-energy gaps [see Figs.~\ref{Fig:Intro}(c) and \ref{Fig:SDW}(a)] can be understood intuitively by the band folding due to $V(x)/2$ and second order perturbation theory in $V(x)/2$.
Note that the gap opening condition corresponds to $4k_F-mQ=0$ with an integer $m$, which is consistent with the commensurate condition discussed previously.

\begin{figure}[t]
	\includegraphics[width=0.4\textwidth]{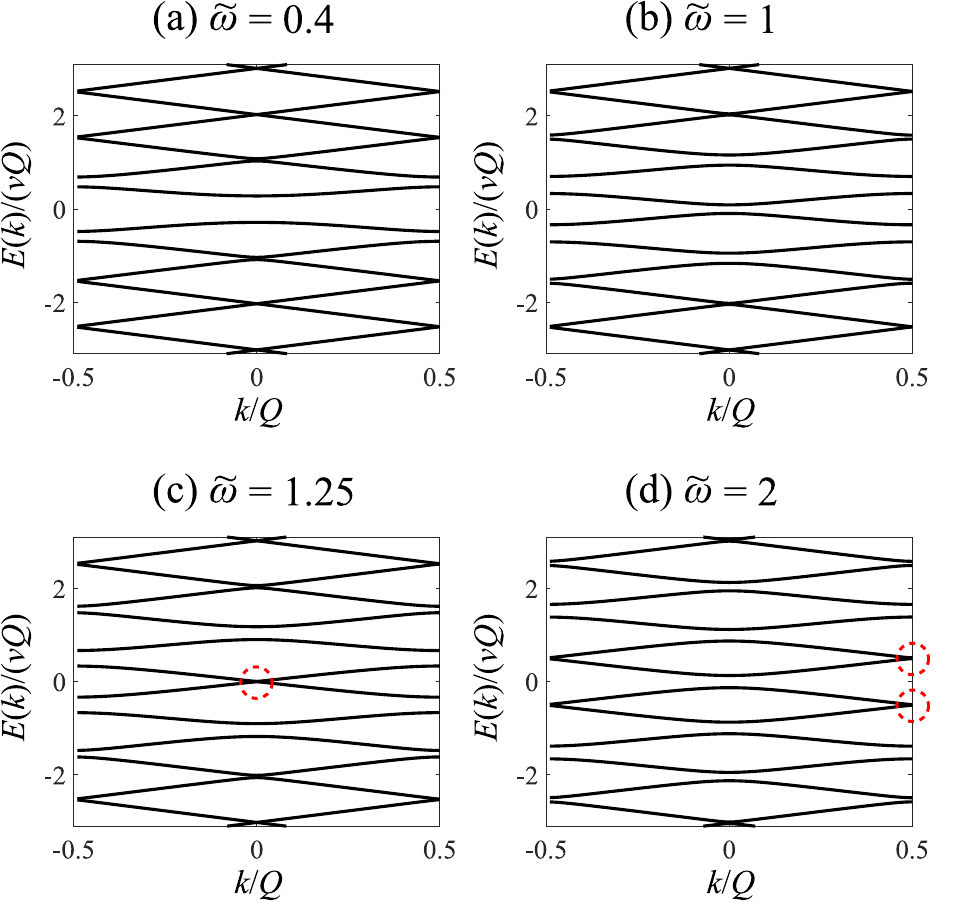}
	\caption{Luther-Emery energy bands for $vQ=3M$, varying $\tilde{\omega}=Kw/(vQ)$ (with $K=1/4$).
	(a) $\tilde{\omega}=0.4$ (b) $\tilde{\omega}=1$ (c) $\tilde{\omega}=1.25$ (d) $\tilde{\omega}=2$. 
	The Luther-Emery bands contain gaps due to spontaneous time-reversal symmetry breaking.
	The Dirac points at $k=0$ and $k=\pm Q/2$ (circled by red dashed lines) are recovered for particular values of $\tilde{\omega}$ in (c) and (d). 
	}
	\label{Fig:LE_band1}
\end{figure}

In addition, controlling the magnitude of the periodic potential can close/reopen the zero-energy and finite-energy gaps.
To see this analytically, we treat the periodic potential as a perturbation and compute
the renormalized velocity and effective mass, which we find to be given by
\begin{align}
	\tilde{v}=vZ\left[1-\alpha^2+\alpha^2\frac{2v^2Q^2}{v^2Q^2+M^2}\right],\,\tilde{M}=MZ\left(1-\alpha^2\right),
\end{align}
respectively, where $Z=1/(1+\alpha^2)$
is the wavefunction renormalization, 
and we have introduced a dimensionless coupling constant $\alpha^2=\frac{w^2}{8\left(v^2Q^2+M^2\right)}$.
Remarkably, the Dirac mass can be renormalized to zero (i.e. $\tilde M=0$) when $\alpha=1$ , suggesting a Dirac point at $E=0$ and band inversion signalled by the change in sign of $\tilde M$.
This is rather interesting as it implies not only can the Dirac point re-emerge but  this process has the potential to  drive topological transitions on the edge.
We confirm the perturbative analysis by numerically solving Eq.~(\ref{Eq:LE_Th}), shown
in Fig.~\ref{Fig:LE_band1}, where we plot the energy bands for a few representative values of $w$, demonstrating the re-emergence of Dirac points at zero energy [Fig.~\ref{Fig:LE_band1}(c)] and finite energies [Fig.~\ref{Fig:LE_band1}(d)].

\begin{figure}[t]
	\includegraphics[width=0.4\textwidth]{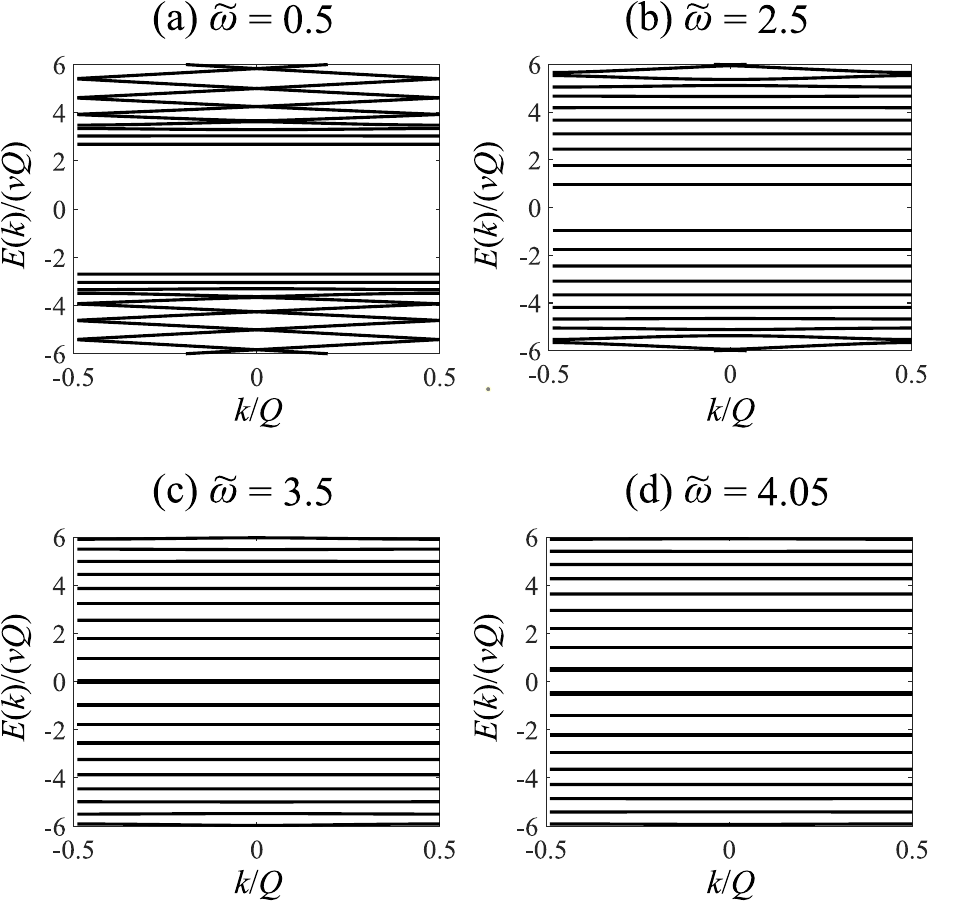}
	\caption{Luther-Emery energy bands for $3vQ=M$, varying $\tilde{\omega}=Kw/(vQ)$ (with $K=1/4$).
	(a) $\tilde{\omega}=0.5$ (b) $\tilde{\omega}=1$ (c) $\tilde{\omega}=3.5$ (d) $\tilde{\omega}=4.05$. 
	The energy bands are nearly flat.
	The gap size at $E=0$ depends on $\tilde{w}$. Nearly doubly degenerate zero energy bands emerge in (c). Similar nearly degenerate bands can also happen at finite energies as shown in (d).}
	\label{Fig:LE_band2}
\end{figure}

For $vQ<|M|$, the low-energy bands can become nearly flat.  
Because of the narrow bandwidth in the low-energy minibands, perturbation theory cannot easily capture the qualitative features in this limit. We present numerical results in Fig.~\ref{Fig:LE_band2}.
Interestingly, we still find the gap closing for particular values of $w$ as shown in Figs.~\ref{Fig:LE_band2}(c) and (d). 
In summary, the periodic potential provides a nontrivial way to manipulate the edge bands, giving rise to finite-energy gaps, flat bands, and re-emergent Dirac points.

\textit{Implication for experiments. --} Our theory describes an unprecedented way to manipulate the emergent spin texture and edge state dispersion by an external periodic potential, $V(x)=w\cos(Qx)$, for the strongly interacting TI edges state ($K<1/2$). The existence of novel SDW states is a manifestation of the spin-momentum locking in the helical edges. 
Based on our prediction, the spin texture can be controlled by both $Q$ and $w$, and the predicted results in Eqs. (\ref{Eq:s_x}) and (\ref{Eq:s_y}) can be tested by scanning SQUID experiments on quantum wells \cite{Nowack2013,Spanton2014}. 

Our results also predict that the edge states can be driven through their own topological transition due to a band inversion. The stability of the bulk gap as the topological edge states are manipulated in such a fundamental way can be investigated using an ultracold Fermi gas with a two-dimensional spin orbit coupling, where both the stability of the bulk gap and the sensitivity to the edge gap can be seen through  radiofrequency spectroscopy measurements of the spectral function ~\cite{gaebler2010observation,PhysRevLett.120.060404}.


Now, we discuss how our setup can enhance edge state coherence in  strongly interacting TI materials.
A TI edge state with $K<1/2$ can develop a gap at $E=0$ due to spontaneous time-reversal symmetry breaking \cite{Wu2006,Xu2006}, causing the edge state to become insulating at $E=0$. For finite energies away from the interacting gap, chemical potential disorder generically induces localization in a 1D massive Dirac model \cite{Bocquet1999,Chou2018}, destabilizing the quantized edge state conduction \cite{Du2015,Li2015,Li2017,Bubis2021localization}. Our work provides a possible resolution to such an almost inevitable scenario for strongly interacting TI edge states. As shown in Fig.~\ref{Fig:LE_band1} and Fig.~\ref{Fig:LE_band2}, in the presence of a periodic potential, the magnitude of the gap depends on both $Q$ and $w$; for appropriate parameters, the gap can be completely suppressed, revealing a re-emergent Dirac point.
Although such parameters are finely tuned, we have shown that in general, the interacting gap at $E=0$ is reduced in the presence of the potential.
Since the transport coherence length (defined by the length scale at which dissipationless edge states break down) increases as the gap decreases, this implies that
 the coherence length of strongly interacting helical edge states can be enhanced by an external periodic potential. 
When the coherence length is much larger than the system size, nearly quantized transport will result.
 

Thus, our work implies that a periodic potential at the edge of a 2D TI can lead to quantized transport even in the presence of interactions.
Our proposal does not require precise information about the local potential landscape, in contrast to the gate training technique used in weakly interacting TI systems \cite{Lunczer2019}.
Since our theory describes interaction-driven edge phases, it is particularly relevant to 2D TI materials exhibiting strongly interacting edge states \cite{Du2015,Li2015,Du2017,Li2017,Bubis2021localization}. 

\begin{acknowledgments} 
This work is supported by the Laboratory for Physical Sciences (Y.-Z.C.), by JQI-NSF-PFC (supported by NSF grant PHY-1607611, Y.-Z.C.).
J.C. acknowledges the support of the Flatiron Institute and the Air Force Office of Scientific Research under Grant No.~FA9550-20-1-0260.
J.H.P is partially supported by  the Air Force Office of Scientific Research under Grant No.~FA9550-20-1-0136.
The Flatiron Institute is a division of the Simons Foundation.
\end{acknowledgments}


\begin{thebibliography}{102}
\bibitem[{U_t()}]{U_term}
	\bibinfo{note}{The minus sign in front of the $U$ term is due to the
		bosonization convention used in this work \cite{Shankar_Book}.}
\bibitem[{FT()}]{FT}
	\bibinfo{note}{As we can see in the Taylor expansion of
		$\exp\left[i\frac{4Kw}{vQ}\sin(Qx)\right]$, the Fourier spectrum is discrete
		with nonzero values at integer multiple of $Q$.}
\end{thebibliography}



\newpage \clearpage 

\onecolumngrid

\begin{center}
	{\large
		Band manipulation and spin texture in interacting moir\'e  helical edges
		\vspace{4pt}
		\\
		SUPPLEMENTAL MATERIAL
	}
\end{center}

\setcounter{figure}{0}
\renewcommand{\thefigure}{S\arabic{figure}}
\setcounter{equation}{0}
\renewcommand{\theequation}{S\arabic{equation}}

In this supplemental material, we provide some technical details of main results in the main text.

\section{Bosonization convention}

We adopt the field-theoretic bosonization convention. The fermionic fields can be described
by chiral bosons via
\begin{align}
	\psi_R(x)=\frac{1}{\sqrt{2\pi a}}e^{i\left[\phi(x)+\theta(x)\right]},\,\,\psi_L(x)=\frac{1}{\sqrt{2\pi a}}e^{i\left[\phi(x)-\theta(x)\right]},
\end{align}
where $\phi$ is the phase-like bosonic field, $\theta$ is the phonon-like bosonic field, and  $a$ is the ultraviolet length scale that is determined by
the microscopic model. The density operator is bosonized into $\hat\rho=\frac{1}{\pi}\partial_x\theta$
The bosons obey $\left[\phi(x),\theta(y)\right]=-i\pi u(y-x)$, where $u$ is the Heaviside function.
We note that Klein factors are not needed with this bosonic commutation relation.
The time-reversal operation
($\mathcal{T}^2=-1$) in the bosonic language is defined as follows:
$\phi\rightarrow -\phi+\frac{\pi}{2}$,
$\theta\rightarrow\theta-\frac{\pi}{2}$, and
$i\rightarrow-i$. This corresponds to the fermionic operation:
$R\rightarrow L$, $L\rightarrow -R$, and $i\rightarrow -i$. 

\section{Bosonized Hamiltonian}

The interacting helical fermions can be mapped to an interacting bosonic problem, describing the low-energy charge collective modes of the 1D linearly dispersing fermions. The bosonized Hamiltonian (corresponding to $\hat{H}$) is given by:
\begin{align}
	\label{Eq:H_b}\hat{H}_{b}=&\int dx \frac{v}{2\pi}\left[K\left(\partial_x\phi\right)^2+\frac{1}{K}\left(\partial_x\theta\right)^2\right]+\frac{w}{\pi}\int dx \cos\left(Qx\right)(\partial_x\theta)-\tilde{U}\int dx\cos\left(4\theta+4k_Fx\right),
\end{align}
where $v$ is the velocity of boson, $K$ is the Luttinger parameter, $\theta$ ($\phi$) is the phonon-like (phase-like) bosonic field, $\tilde{U}=U/(2\pi^2\alpha^2)$, and $\alpha$ is the ultraviolet length scale in the low-energy theory. Note that the minus sign in front of the $U$ term is due to the bosonization convention used in this work.
The density and current operators can be expressed by $\hat\rho=\partial_x\theta/\pi$ and $\hat{I}=-\partial_t\theta/\pi$ respectively. We focus only on repulsive interactions, corresponding to $K<1$.

\section{Spin density bilinears}

The spin density operators are given by
\begin{align}
	s_x(x)\!=&e^{i2k_Fx}\psi_L^{\dagger}\psi_R(x)\!+\!\text{H.c.}\!\rightarrow\!\frac{1}{\pi a}\!\sin\left[2\theta(x)\!+\!2k_Fx\right],\\
	s_y(x)\!=&ie^{i2k_Fx}\psi_L^{\dagger}\psi_R(x)\!+\!\text{H.c.}\!\rightarrow\!\frac{1}{\pi a}\!\cos\left[2\theta(x)\!+\!2k_Fx\right],
\end{align}
where $s_x$ and $s_y$ correspond to the $x$ and $y$ component of the spin density operators respectively. 

\section{Charge ordering pattern}

The spin density wave state can also be viewed as a charge density wave state due to the ``spin-momentum locking'' of the helical edge fermions. To see this, recall that the density operator is expressed by $\rho=\partial_x\theta/\pi$. Employing a semiclassical analysis (valid for $K\ll 1$) by ignoring the quadratic boson terms in Eq.~(\ref{Eq:H_b}), the semiclassical Hamiltonian with $k_F=0$ is given by
\begin{align}\label{Eq:H_cl}
	\hat{H}_{\text{cl}}=\frac{w}{\pi}\int dx \cos\left(Qx\right)(\partial_x\theta)-\tilde{U}\int dx\cos\left(4\theta\right).
\end{align}
The second term is minimized when $\theta=\pi P/2$ where $P$ is an integer, while the first term modulates $\theta$ in space. As a result of this competition, localized charge puddles ($\partial_x\theta/\pi>0$) and hole puddles ($\partial_x\theta/\pi<0$) develop for a sufficiently large $w$. A more careful analysis \cite{Chou2018} reveals that the puddles are made of $\pm e/2$ solitons(anti-solitons).
For $k_F\neq0$, similar conclusions apply except that $\int dx\rho(x)\neq 0$ implies different numbers of electrons and holes.
Therefore, the spin texture of a SDW state is associated with a charge pattern with electron-like and hole-like localized ``puddles.'' Within the semiclassical analysis, a smaller $Q$ implies a larger inter-puddle distance. The value of $w$ controls the number of solitons (anti-solitons) in each puddle.
Intuitively, the insulating state might be destroyed for a sufficiently smalle $Q$, indicating the hybridization between localized puddles. 
On the contrary, for a sufficiently large $Q$, localized charge puddles that are well-separated (i.e., inter-puddle hybridization can be ignored), and the system can be described by asymptotically localized orbitals, corresponding to a featureless flat dispersion. These intuitions are backed up by the exact Luther-Emery calculations in the main text.

\section{Perturbative analysis for Luther-Emery Green function}

In this section, we perform a perturbative analysis by treating $V(x)$ as a small perturbation. We focus on only the low-energy Green function which encodes essential information such as the gap at $E=0$ and velocity.

The Luther-Emery Hamiltonian can be expressed by the first quantized Hamiltonian $\hat{h}_0+\hat{V}$,
where
\begin{align}
	\hat{h}_0=&v\hat{\sigma}_z\left(-i\partial_x\right)-M\hat{\sigma}_y,\\
	\hat{V}=&\frac{w}{4}\left[e^{iQx}+e^{-iQx}\right]\hat{1}.
\end{align}
In the above expressions, $\sigma_{a}$ is the $a$-component of the Pauli matrix and $\hat{1}$ is the $2\times 2$ identity matrix. Note that the mass term is $-M$.
The Green function of $\hat{h}_0+\hat{V}$ can be constructed formally by
\begin{align}
	\label{Eq:G_expansion}	\hat{G}\equiv&\frac{1}{E\hat{1}-\hat{h}_0-\hat{V}}=\frac{1}{\hat{G}_0^{-1}-\hat{V}}=\hat{G}_0+\hat{G}_0\hat{V}\hat{G}_0+\hat{G}_0\hat{V}\hat{G}_0\hat{V}\hat{G}_0+\dots,
\end{align}
where $\omega$ is the frequency and 
\begin{align}
	\hat{G}_0(E,k)=\frac{1}{\omega \hat{1}-\hat{h}_0(k)}=\frac{E\hat{1}+vk\hat{\sigma}_z-M\hat{\sigma}_y}{E^2-(vk)^2-M^2}.
\end{align}	

We treat $\hat{V}$ as perturbation and construct a self energy $\hat{\sigma}$ at order $w^2$ as follows:
\begin{align}
	\hat{\Sigma}(E,k)=\left(\frac{w}{4}\right)^2\left[\frac{E\hat{1}+v(k+Q)\hat{\sigma}_z-M\hat{\sigma}_y}{E^2-v^2(k+Q)^2-M^2}+\frac{E\hat{1}+v(k-Q)\hat{\sigma}_z-M\hat{\sigma}_y}{E^2-v^2(k-Q)^2-M^2}
	\right].
\end{align}
In the limit $E,vk\ll vQ,M$, we can derive an asymptotic expression for $\hat{\Sigma}$, given by
\begin{align}
	\hat{\Sigma}(E,k)\approx -\left[\delta h E\hat{1}+\delta v k\hat{\sigma}_z-\delta M\hat{\sigma}_y\right],
\end{align}
where
\begin{align}
	\delta h=&\frac{w^2}{8\left(v^2Q^2+M^2\right)}\equiv \alpha^2,\\
	\delta v=&\alpha^2 v\left[1-\frac{2v^2Q^2}{v^2Q^2+M^2}\right],\\
	\delta M=&\alpha^2 M.
\end{align}
The Green function is approximated by
\begin{align}
	\hat{G}\approx\frac{1}{\hat{G}^{-1}(E,k)-\hat{\Sigma}}
	=\frac{Z}{E\hat{1}-\tilde{v}k\hat{\sigma}_z+\tilde{M}\hat{\sigma}_y},
\end{align}	
where
\begin{align}
	Z=&\frac{1}{1+\alpha^2},\\
	\tilde{v}=&v\frac{1}{1+\alpha^2}\left[1-\alpha^2+\alpha^2\frac{2v^2Q^2}{v^2Q^2+M^2}\right],\\
	\tilde{M}=&M\frac{1-\alpha^2}{1+\alpha^2}.
\end{align}

The results of the dressed Green function suggest that (a) $\tilde{v}$ can vanish if $v^2Q^2<M^2$, and (b) $\tilde{M}=0$ at $\alpha=1$. The most striking aspect is that the gap at $E=0$ vanishes (i.e. $\tilde{M}=0$) for a fine-tuned value of $w$. The vanishing of velocity indicates that the linear-in-momentum coefficient is zero in the dispersion. We should expect higher order terms that might change the predictions based on leading order perturbation theory. 

Some of the results here can be confirmed numerically. In particular, we find that the gap can vanish but the magic value of $w$ is different from the perturbative analysis. We also find that the velocity can become very small, i.e., bandwidth is very narrow. However, we do not find velocity inversion (i.e., negative velocity) in the numerics.

\section{Luther-Emery band structure calculations}

The Luther-Emery theory in the main text can be diagonalized numerically. The Dirac equation is given by
\begin{align}\label{Eq:Dirac_E}
	\left[\hat{h}_0+\hat{V}\right]\Psi_k(x)=E\Psi_k(x),
\end{align}
where $\Psi_k^T=[\tilde{\Psi}_R(k), \tilde{\Psi}_L(k)]$ is the eigen wavefunction with the momentum $k$. We note that the system is periodic when $x\rightarrow x+\frac{2\pi}{Q}$. Thus, we can express the $\Psi_k(x)$ by
\begin{align}\label{Eq:Psi}
	\Psi_k(x)=&\sum_{s=\pm}\sum_m C_{m,s}|k+mQ,s\rangle,\\
	\langle x|k+mQ,s\rangle=&\hat{\Phi}_s(k+mQ)e^{i(k+mQ)x},
\end{align}
where $m$ runs over all the integer, $s$ is the index of the positive and negative energy band, $\hat{\Phi}_s(k)$ is the eigen-spinor of $\hat{h}_0$, and $C_{m,s}$ is the expansion coefficient. The eigen-spinors are given by
\begin{align}
	\hat{\Phi}_+(k)=&\frac{1}{\sqrt{2\left(v^2k^2+M^2\right)+2vk\sqrt{v^2k^2+M^2}}}\left[\begin{array}{c}
		vk+\sqrt{v^2k^2+M^2}\\
		-iM
	\end{array}\right],\\
	\hat{\Phi}_-(k)=&\frac{1}{\sqrt{2\left(v^2k^2+M^2\right)-2vk\sqrt{v^2k^2+M^2}}}\left[\begin{array}{c}
		-vk+\sqrt{v^2k^2+M^2}\\
		iM
	\end{array}\right],
\end{align}	
corresponding to positive ($E_{0+}=\sqrt{v^2k^2+M^2}$) and negative ($E_{0-}=-\sqrt{v^2k^2+M^2}$) energies respectively.
Note that the inner product of the spinors at the same momentum, $\hat{\Phi}_{s}^{\dagger}(k)\hat{\Phi}_{s'}(k)=\delta_{s,s'}$. However, $\hat{\Phi}_{s}^{\dagger}(k)\hat{\Phi}_{s'}(k')$ is generically nonzero.

With Eq.~(\ref{Eq:Psi}), the eigenvalue problem in Eq.~(\ref{Eq:Dirac_E}) can be expressed by
\begin{align}
	&\sum_{m,s}\langle k+nQ,s' |\left[\hat{h}_0+\hat{V}\right]C_{m,s}|k+mQ,s\rangle=EC_{n,s'},\\
	\label{Eq:Band_diag}\rightarrow &s'\sqrt{v^2(k+nQ)^2+M^2}C_{n,s'}+\frac{w}{4}\sum_{m,s}C_{m,s}\left[\mathcal{V}^{s's}_{m+1,m}(k)+\mathcal{V}^{s's}_{m-1,m}(k)\right]=EC_{n,s'},
\end{align}	
where
\begin{align}
	\mathcal{V}^{s's}_{n,m}(k)=\hat{\Phi}_{s'}^{\dagger}(k+nQ)\hat{\Phi}_{s}(k+mQ).
\end{align}

Equation~(\ref{Eq:Band_diag}) is infinite dimensional. However, we can impose a truncation in $m$ as we are only interested in bands near $E=0$. For definiteness, we consider $m$ sums over from $-N$ to $N$, where $N$ is a positive integer. The energy cutoff is approximated $\Lambda=vNQ$. The truncation is valid as long as $M,w,E\ll\Lambda$. Numerically, one can vary the value of $N$ to determine if the bands converge in the target energy range.

\end{document}